\documentclass[11pt]{article}
\usepackage{amsfonts,amssymb,amsmath}

\def\smallspace{\mskip 2mu minus 1mu}
\def\tinyspace{\mskip 1mu}

\def\negtinyspace{\mskip -1mu}

\newcommand{\vph}[2]{\vphantom{#1}\smash{#2}}

\def\squareforqed{\hbox{\rlap{$\sqcap$}$\sqcup$}}
\def\qed{\ifmmode\squareforqed\else{\unskip\nobreak\hfil
\penalty50\hskip1em\null\nobreak\hfil\squareforqed
\parfillskip=0pt\finalhyphendemerits=0\endgraf}\fi}
\newtheorem{theorem}{Theorem}
\newtheorem{lemma}[theorem]{Lemma}
\newtheorem{corollary}[theorem]{Corollary}

\newenvironment{proof}{\begin{trivlist}\item[]{\flushleft\bf Proof }}
{\qed\end{trivlist}}
\newcommand{\pftext}[1]{{\bf #1 }}

\newcommand{\ket}[1]{\mbox{$| #1 \rangle$}}

\newcommand{\inner}[2]{\mbox{$\langle #1 | #2 \rangle$}}

\DeclareMathSymbol{\itTheta}{\mathalpha}{letters}{"02}
\DeclareMathSymbol{\leqslant}{\mathrel}{AMSa}{"36}  
\newcommand{\subgroup}{\leqslant}        %subgroup-symbol
\newcommand{\isequal}{\vph{=}{\stackrel{{\scriptscriptstyle ?}}{=}}}
\newcommand{\sqrtinv}[1]{\frac{\smallspace 1}{\sqrt{#1\tinyspace}\tinyspace}}

\newcommand{\integer}{{\mathbb Z}}       % the integers
\newcommand{\Co}{{\mathbb C}}            % the field of complex numbers
\newcommand{\func}{\gamma}
\newcommand{\viola}{\mbox{\boldmath$\mathcal V$}} 
\newcommand{\rv}[1]{\mbox{$\mathbf #1$}}

\title{On Quantum Algorithms for\\Noncommutative Hidden Subgroups}

\author{Mark Ettinger\,$^{1,}$\thanks{\,Email: 
\texttt{\boldmath ettinger$\mathchar"40$lanl.gov}.}
\qquad \qquad
Peter H{\o}yer\,$^{2,1,}$\thanks{\,Email: 
\texttt{\boldmath u2pi$\mathchar"40$imada.ou.dk}.}\vspace{.4cm}\\
{\protect\small $^1$\,\sl Los Alamos National Laboratory}\\
{\protect\small $^2$\,\sl Odense University}}
\date{}

\begin{document}
\maketitle

Report No.~LA-UR-98-2010 (May~6, 1998) \vspace{.6cm}

\begin{abstract}
Quantum algorithms for factoring and discrete logarithm
have previously been generalized to 
finding hidden subgroups of finite Abelian groups.
This paper explores the possibility of extending this general viewpoint to
finding hidden subgroups of noncommutative groups.  
We~present a quantum algorithm for the special case of dihedral groups 
which determines the hidden subgroup 
in a linear number of calls to the input function.  
We~also explore the difficulties of developing an algorithm to process the
data to explicitly calculate a generating set for the subgroup.  
A~general framework for the noncommutative hidden subgroup problem 
is discussed and we indicate future research directions.
\end{abstract}

\section{Introduction}\label{sec:intro}
All known quantum algorithms which run super-polynomially faster than 
the most efficient probabilistic classical algorithm
solve special cases of what is called
the Abelian Hidden Subgroup Problem.  
This general formulation includes Shor's algorithms 
for factoring and finding discrete logarithms~\cite{Shor97}.
A~very natural question to ask is if quantum computers can efficiently
solve the Hidden Subgroup Problem in {\em noncommutative\/} groups.  
This question has been raised 
regularly~\cite{Beals97,Hoyer97,Jozsa98,Kitaev95}, 
and seems important for at least three reasons. 
\looseness-1

The first reason is that determining if two graphs are isomorphic 
reduces to finding hidden subgroups of symmetric groups.  
The second reason is that the noncommutative hidden subgroup problem 
arguably represents a most natural line of research 
in the area of quantum algorithmics.
The third reason is that an efficient quantum algorithm for 
a hidden subgroup problem could potentially be used to show an 
exponential gap between quantum and classical two-party probabilistic
communication complexity models~\cite{CB97,BCW98}.

The heart of the idea behind the quantum 
solution to the Abelian hidden subgroup problem
is Fourier analysis on Abelian groups.
The difficulties of Fourier analysis on noncommutative groups 
makes the noncommutative version of the problem very challenging.  

In~this paper, we present the first known quantum algorithm for a
noncommutative subgroup problem.
We~focus on dihedral groups because they are 
well-structured noncommutative groups, and because they contain an 
exponentially large number of different subgroups of small order,
making classical guessing infeasible.
Our main result is that there exists a quantum algorithm 
that solves the dihedral subgroup problem using only 
a linear number of evaluations of the function which is given as input.  
This is the first time such a result has been obtained for
a noncommutative group.  
However, we hasten to add that our algorithm does {\em not\/} run in 
polynomial time, even though it only uses few evaluations of the 
given function.
The reason for this is as follows:
The algorithm applies a certain quantum subroutine 
a linear number of times, each time producing some output data.
The collection of all the output data 
determines the hidden subgroup with high probability.
We~know how to find the subgroup from the data in exponential time,
but we do not know if this task can be done efficiently.

Three important questions are left open.  The first question is if there
exists a polynomial-time algorithm (classical or quantum) to postprocess
the output data from our quantum subroutine. 
The second is whether our algorithm can be used to show an exponential
gap between quantum and classical probabilistic
communication complexity models, as mentioned above.
Currently, the state-of-the-art is an exponential separation
between error-free models, and a quadratic separation
between probabilistic models~\cite{BCW98}.
The third open question is for what other noncommutative groups 
similar results can be obtained.

\section{Algorithm for dihedral groups}\label{sec:alg}
The {\em Hidden Subgroup Problem\/} is defined as follows:
{\begin{description}
\itemsep0cm
\item[{\bf Given}:]  A~function $\func: G \rightarrow R$, where
  $G$ is a finite group and $R$ an arbitrary finite range.
\item[{\bf Promise}:]  There exists a subgroup $H \subgroup G$
  such that $\func$ is {\em constant\/} and {\em distinct\/} on 
  the left cosets of~$H$.
\item[{\bf Problem}:]  Find a generating set for~$H$.
\end{description}}
We say of such a function~$\func$ that it {\em fulfills the subgroup
promise\/} with respect to~$H$.  
We also say of~$\func$ that it {\em has hidden subgroup~$H$}.
Note that we are not given the order of~$H$. 
Without loss of generality we assume $\func$ is constant and distinct on 
{\em left\/} cosets because we may formally rename group elements and convert 
multiplication on the right to multiplication on the left.

If $G$ is Abelian, then we refer to this problem as the
Abelian Subgroup Problem.
Similarly, if the given group is dihedral, then we 
refer to it as the Dihedral Subgroup Problem.
Classically, if $\func$ is given as a black~box, 
then the Abelian subgroup problem is infeasible:
If $G = \integer_2^n$, then 
just to determine if $H$ is non-trivial or not takes time exponential 
in~$n$~\cite{Simon97}.
Here, $\integer_2$ denotes the cyclic group of order~2.
In~contrast, the Abelian subgroup problem can be 
solved efficiently on a quantum 
computer~\cite{BL95,BH97,Grigoriev97,Kitaev95,Shor97,Simon97}.

\begin{theorem}\label{thm:abelian}
Let $\func : G \rightarrow R$ be a function that fulfills
the {\em \/Abelian} subgroup promise with respect to~$H$.
There exists a quantum algorithm that outputs a subset~$X \subseteq G$
such that $X$ is a generating set for~$H$ 
with probability at least~\mbox{$1-1/|G|$},
where $|G|$ denotes the order of~$G$.
The algorithm uses $O(\log |G|)$ evaluations of~$\func$,
and runs in time polynomial in~$\log |G|$ and in the time 
required to compute~$\func$.
\end{theorem}

We~review the quantum solution to the Abelian subgroup problem 	
in terms of group representation theory
in Section~\ref{sec:abelian} below.
For other reviews, see for example~\cite{BH96,Jozsa98}.

The dihedral group of order~$2N$ is the symmetry group of an
$N$--sided polygon.  
It~is a semidirect product of the two cyclic groups $\integer_N$ 
and~$\integer_2$ of order $N$ and~2, respectively.
It~is isomorphic to the group
\begin{equation}\label{eq:dihedral}
D_N = \integer_N \rtimes_\phi \integer_2
\end{equation}
with the multiplication defined by
\[(a_1,b_1)(a_2,b_2) 
  = \big(a_1 + \phi(b_1)(a_2), \smallspace b_1 + b_2\big),\]
where the homomorphism 
$\phi : \integer_2 \rightarrow \textup{Aut}(\integer_N)$
is defined by \mbox{$1 \mapsto \phi(1)(a)  = -a$}.

\begin{theorem}[Main theorem]\label{thm:general}
Let \mbox{$\func : D_N \rightarrow R$} be a function that 
fulfills the {\em \/dihedral} subgroup promise with respect to~$H$.
There exists a quantum algorithm that given~$\func$,
uses $\itTheta(\log N)$ evaluations of~$\func$ and
outputs a subset $X \subseteq D_N$ such that $X$ is a 
generating set for~$H$ with probability at least \mbox{$1-\frac2{N}$}. 
\end{theorem}

Theorem~\ref{thm:general} constitutes our main result that 
the dihedral subgroup problem can be solved with few applications
of the given function~$\func$.
The essential step in the proof is that it is possible to
find subgroups of order~$2$.  
The dihedral group~$D_N$ contains $N+1$~different subgroups of 
order~2 if $N$ is even, 
and $N$ different subgroups of order~$2$ if $N$ is odd.

So, even if we are promised that the hidden subgroup
is of that order, a straight-forward approach to find its generator
would take time exponential in~$\log(N)$.
Theorem~\ref{thm:special} entails that we can find
the generator with an expected number of evaluations of~$\func$ only
linear in~$\log N$.

\begin{theorem}\label{thm:special}
Let \mbox{$\func : D_N \rightarrow R$} be a function that 
fulfills the dihedral subgroup promise with respect to~$H$, 
where $H$ is either the trivial subgroup, or $H=\{(0,0),(k_0,1)\}$
for some $0 \leq k_0 < N$.
There exists a quantum algorithm that given $\func$,
uses at most $89 \log(N) +7$ evaluations of~$\func$ 
and outputs either ``trivial'' or the value~$k_0$.
If~$H$ is trivial then the output is always ``trivial'', and
if~$H$ is non-trivial then the algorithm outputs~$k_0$ 
with probability at least~$1-\frac{1}{2N}$.
\end{theorem}

We~first give the reduction of the general problem given in
Theorem~\ref{thm:general} to the special case in Theorem~\ref{thm:special}.

\begin{proof}\pftext{of Theorem~\ref{thm:general}}
Let $\func_1$ denote the restriction of $\func$ to
the cyclic subgroup $\integer_N \times \{0\} \subgroup D_N$ of order~$N$.
Then $\func_1 : \integer_N \times \{0\} \rightarrow R$ fulfills the
Abelian subgroup promise with respect 
to \mbox{$H_1 = H \cap (\integer_N \times \{0\})$}.
By~Theorem~\ref{thm:abelian}, we can,
by using $O(\log N)$ evaluations of~$\func_1$, 
find a subset $X_1 \subseteq H_1$ so that $X_1$ generates~$H_1$
with probability at least~$1-1/N$.

The subgroup $\langle X_1 \rangle \subgroup D_N$ is normal in~$D_N$,
and the factor group $D_N / \langle X_1 \rangle$ is isomorphic 
to~$D_M$ where 
$M = \min\{1 \leq j \leq N \mid (j,0) \in \langle X_1 \rangle\}$.
Since $\func$ is constant on the cosets of $\langle X_1 \rangle$,
we can consider $\func$ a function~$\func_2$ on~$D_M$.
Then $\func_2 : D_M \rightarrow R$ fulfills the dihedral subgroup 
promise with respect to some subgroup~$H_2 \subgroup D_M$.

Suppose $\langle X_1 \rangle = H_1$.  
Then either $H_2 = \{(0,0)\}$ is the trivial subgroup,
or $H_2 = \{(0,0),(k_0,1)\}$ for some $0 \leq k_0 < M$.
Further, if $H_2$ is trivial then $H = \langle X_1 \rangle$,
and if $H_2 = \{(0,0),(k_0,1)\}$ then $H = \langle X_1, (k_0,1) \rangle$.

We~now apply the algorithm in Theorem~\ref{thm:special} 
with~$\func_2 : D_M \rightarrow R$, 
producing either ``trivial'' or~$k_0$.  We~repeat this 
$t = \raisebox{.2mm}{$\big\lceil$}\!\log(2N)/\log(2M) 
     \negtinyspace\raisebox{.2mm}{$\big\rceil$}$
times in total, ensuring we will find~$k_0$ with probability 
at least~$1 - 1/(2M)^t \geq 1 - 1/2N$, provided $k_0$ exists.
If~we obtain~$k_0$, then let $X= X_1 \cup \{(k_0,1)\}$, 
and otherwise let $X=X_1$.

If~$X_1$ generates $H_1$, then with
probability at least~$1-1/2N$ we have $H = \langle X \rangle$.
Since $X_1$ generates $H_1$ with probability at least~\mbox{$1-1/N$},
the overall success probability is at least 
\mbox{$(1-1/N)(1-1/2N) > 1-2/N$}.
The total number of evaluations of~$\func$ is 
at most $O(\log N) + t (89 \log M + 7)$,
as each evaluation of~$\func_1$ and~$\func_2$
requires one evaluation of~$\func$.
\end{proof}

We~assume that the reader is familiar with the basic notions
of quantum computation~\cite{Berthiaume97}.
The quantum algorithm we shall use to prove 
Theorem~\ref{thm:special}~is
\begin{equation}\label{eq:viola}
{\viola}_\func \;=\;
  \big({\mathbf F}_N \otimes {\mathbf W} \otimes {\mathbf I}\big) 
  \,\circ\, {\mathbf U}_\func \,\circ\,
  \big({\mathbf F}_N^{-1} \otimes {\mathbf W} \otimes {\mathbf I}\big).
\end{equation}
Here, ${\mathbf U}_\func$ is any unitary operator that satisfies that
\begin{equation}\label{eq:Ufunc}
{\mathbf U}_\func \;\ket{a}\ket{b}\ket{0} = \ket{a}\ket{b}\ket{\func(a,b)}
\end{equation}
for all $0 \leq a < N$ and $0 \leq b \leq 1$.
The operator~${\mathbf F}_N$ is the quantum Fourier transform 
for~$\integer_N$ defined~by
\begin{equation}
{\mathbf F}_N \,\ket{i} \;=\; \sqrtinv{N}
  \sum_{j=0}^{N-1} \omega^{i j}_N \;\ket{j},
\end{equation}
where $\omega_N = e^{2 \pi \sqrt{-1}/N}$ is the $N$th principal root
of unity.
When~$N=2$, then the Fourier transform~${\mathbf F}_2$ 
is equal to the Walsh--Hadamard transform~${\mathbf W}$ 
which maps a qubit in state~\ket{b} to the superposition 
\mbox{$\sqrtinv{2} \big(\ket{0} + (-1)^b \ket{1}\big)$}.

Suppose for a moment that we were not given a function 
defined on the dihedral group~$D_N = \integer_N \rtimes_\phi \integer_2$,
but instead a function defined 
on the Abelian group~$\integer_N \times \integer_2$.
Or~equivalently, suppose for the moment that 
\mbox{$\phi : \integer_2 \rightarrow \textup{Aut}(\integer_N)$} 
is the trivial homomorphism.
Then by Theorem~\ref{thm:abelian}, we can find any hidden subgroup
with probability exponentially close to~1
by applying the experiment
\begin{equation}\label{eq:experiment}
(a,b) = {\mathcal M}_{1,2}\; \circ {\viola}_\func
          \,\ket{0}\ket{0}\ket{0}
\end{equation}
a number of $O(\log N)$ times.
Here, ${\mathcal M}_{1,2}$ denotes a measurement of the first 
two registers with outcome~$(a,b)$.
A~natural question to ask is, 
how much information, if any,  would we gain by performing the 
experiment given in Equation~\ref{eq:experiment} when $\func$ 
is defined on~$D_N$ and not on~$\integer_N \times \integer_2$.
The next lemma shows that we indeed learn something.

\begin{lemma}\label{lm:sampling}
Let $\func : D_N \rightarrow R$ fulfill the subgroup 
promise with respect to $H = \{(0,0),(k_0,1)\}$.
Then, if~we apply quantum algorithm ${\viola}_\func$ on the initial
state \ket{0}\ket{0}\ket{0}, the probability that the outcome 
of a measurement of the first two registers is $(a,0)$,~is 
\begin{equation}\label{eq:sampling0}
\frac1{2N} \big(1+\cos(2 \pi k_0 a /N)\big) 
  = \frac1N \cos^2(\pi k_0 a/N).
\end{equation}
Furthermore, the probability that the outcome is~$(a,1)$, 
is~$\frac1N \sin^2(\pi k_0 a/N)$.
\end{lemma}

Let $\rv{Z}$ denote the discrete random variable defined by the 
probability mass function
\[\textup{Prob}[\rv{Z}=z] \;=\; \alpha \cos^2(\pi k_0 z/N)
  \qquad (z \in \integer_N),\]
where $\alpha = 1/N$ if $k_0=0$ or $\tinyspace{}2k_0=N$, 
and $\alpha=2/N$ otherwise.
Lemma~\ref{lm:sampling} provides us with a quantum algorithm
for sampling from~\rv{Z}.
Intuitively, since~\rv{Z} is non-uniformly distributed on~$\integer_N$ 
depending on~$k_0$, the more samples we draw from~\rv{Z}, the more
knowledge we gather about~$k_0$.  
The crucial question therefore becomes, how many samples from~\rv{Z}
do we need to be able to identify~$k_0$ correctly with high probability.
Theorem~\ref{thm:peak} below states that we only need a
logarithmic number of samples.  
We~postpone its proof till the next section.

\begin{theorem}\label{thm:peak}
Let $m \geq \raisebox{.35mm}{$\lceil$} 64 \ln N \raisebox{.35mm}{$\rceil$}$, 
and let $z_1,\dots,z_m$ be independent samples from~$\rv{Z}$.
Let ${\tilde k} \in \{1,\dots,\lfloor N/2 \rfloor \}$ be such that 
the sum $\sum_{i=1}^m \cos(2 \pi {\tilde k} z_i/N)$ is maximal.
Then ${\tilde k} = \min\{k_0,N-k_0\}$ 
with probability at least~\mbox{$1-\frac{1}{2N}$}.
\end{theorem}

\begin{proof}\pftext{of Theorem~\ref{thm:special}}
The algorithm starts by disposing the possibility that $k_0=0$ 
by computing $\func(0,0)$ and~$\func(0,1)$.
If~the two values are equal, 
then the algorithm outputs the value~$0$ and stops.
If $N$~is even, then the algorithm proceeds by disposing 
the possibility that $k_0=N/2$, too.

Now, the algorithm applies the quantum experiment given
in Equation~\ref{eq:experiment} a number of 
$m' = 2 \raisebox{.35mm}{$\lceil$} 64 \ln N \raisebox{.35mm}{$\rceil$}$
times.
Let $m$ denote the number of times it measures a~0 in the second register.
Let $\{a_1,\dots,a_m\}$ denote the outcomes in the first register,
conditioned to that the measurement of the second register 
yields~a~zero.\footnote{\;Alternatively, we could apply amplitude 
amplification~\cite{BH97} to ensure that we will always measure~0 
in the second register, instead of as here, only with 
probability~\mbox{1/2}.}

Suppose $m \geq m'/2$.
The algorithm continues with classical post-processing:
It finds $1 \leq \tilde k \leq \lfloor N/2 \rfloor$ such that
the sum $\sum_{i=1}^m \cos(2 \pi {\tilde k} a_i/N)$ is maximized.
It~then computes $\func({\tilde k},1)$ 
and compares it with the previous calculated value~$\func(0,0)$.
If~they are equal, it outputs~$\tilde k$ and stops.
Otherwise, it performs the same test for $\func(N-{\tilde k},1)$.
If that one also fails, it outputs \mbox{``trivial''}.

If $m < m'/2$, then the algorithm performs the same classical 
post-processing,
except that it uses the $m'-m$ measurements for which 
the output in the second register is~1, and except that it now 
seeks to maximize the sum $\sum_{i=1}^m \sin(2 \pi {\tilde k} a_i/N)$.

If $H$ is trivial, then the algorithm returns ``trivial'' with certainty.
If~$H = \{(0,0),(k_0,1)\}$,
then it outputs~$k_0$ with 
probability at least~$1-1/2N$ by Theorem~\ref{thm:peak}.
The total number of evaluations of~$\func$ is upper
bounded by~\mbox{$m'+ 5 < 89 \log N + 7$}.
\end{proof}

\section{Proof of Theorem~\ref{thm:peak}}\label{sec:peak}
The proof of Theorem~\ref{thm:peak} requires two lemmas, 
the first of them being a result by~Hoeffding~\cite{Hoeffding63} 
on the sum of bounded random variables.
Hoeffding's lemma says that the probability that the sum of
$m$~independent samples are off from its expected value by a 
constant fraction in~$m$ drops exponentially in~$m$.

\begin{lemma}[Hoeffding]\label{lm:hoeffding}
Let $\rv{X}_1, \dots, \rv{X}_m$ be independent identically dis\-tri\-buted
random variables with $\ell \leq \rv{X}_1 \leq u$.
Then, for all $\alpha > 0$,
\[\textup{Prob}[\rv{S} - \textup{E}[\rv{S}] \geq \alpha m] 
   \;\leq\; e^{-2 \alpha^2 m/ (u-\ell)^2}\]
where $\rv{S} = \sum_{i=1}^m \rv{X}_i$.
\end{lemma}

Let $0 < k < N$, and suppose we want to test if 
$k~\isequal~k_0$ or \mbox{$k~\isequal~N-k_0$},
where $k_0$ is given as in Lemma~\ref{lm:sampling}.
Clearly, we can answer that question just by testing
if $\func(0,0)~\isequal~\func(k,1)$
or $\func(0,0)~\isequal~\func(N-k,1)$.
Lemma~\ref{lm:singlepoint} provides us with another probabilistic
method:  First draw $m$~samples~$\{z_i\}_{i=1}^m$ from~\rv{Z}, 
and then compute the sum~$\sum_{i=1}^m \cos(2 \pi k z_i/N)$.
Conclude that $k \neq k_0$ and \mbox{$k \neq N-k_0$}
if and only if that sum is at most~$m/4$.

\begin{lemma}\label{lm:singlepoint}
Let $0 < k < N$.
Let $z_1,\dots,z_m$ be $m$ independent samples from~$\rv{Z}$.
Then with probability at most $e^{-m/32}$,
we have 
\[\sum_{i=1}^m \cos(2 \pi k z_i/N) \leq m/4\]
if $k=k_0$ or $k=N-k_0$, and
\[\sum_{i=1}^m \cos(2 \pi k z_i/N) \geq m/4\]
otherwise.
\end{lemma}

\begin{proof}
Let $f$ denote the function of $\rv{Z}$ defined by
$f(z) = \cos(2 \pi k z /N)$,
and let $\rv{X} = f(\rv{Z})$ denote the random variable defined by~$f$.
Then $-1 \leq \rv{X} \leq 1$ and the expected value of~\rv{X} is
%P \begin{align}
\[\textup{E}[\rv{X}] 
%P  &= \sum_{z=0}^{N-1} \cos(2 \pi k z /N) \,\textup{Prob}[\rv{Z}=z]\notag\\
%P  &= \sum_{z=0}^{N-1} \cos(2 \pi k z /N) 
%P          \frac2N \cos^2(\pi k_0 z/N)\notag\\
%P  &
= \begin{cases}
       \;1 \quad &\text{if $2k = 2k_0 = N$}\\
       \;\frac12 \quad &\text{if either $k = k_0$ or $k = N - k_0$}\\
       \;0 \quad &\text{otherwise.}
     \end{cases}\]
%P\end{align}
If~$k \neq k_0$ and $k \neq N-k_0$, then apply
Hoeffding's lemma on $m$~independent random variables all
having the same probability distribution as~\rv{X}.
If~$k = k_0$ or $k = N-k_0$, then apply
Hoeffding's lemma on $m$~independent random variables all
having the same probability distribution 
as the random variable~\mbox{$\textup{E}[\rv{X}] - \rv{X}$}.
\end{proof}

If~we are only concerned about testing for a specific $0 < k < N$
if $k~\isequal~k_0$ or \mbox{$k~\isequal~N-k_0$}, then 
Lemma~\ref{lm:singlepoint} is not beneficial since we could
just test if $\func(0,0)~\isequal~\func(k,1)$
or $\func(0,0)~\isequal~\func(N-k,1)$.
But since we want to test all possible values of~$k$,
and not only a single one, then the method yielded
by Lemma~\ref{lm:singlepoint} becomes valuable,
provided we can reuse the same $m$~samples in all tests.
We~now prove Theorem~\ref{thm:peak} by showing that, 
given a set of $m$~samples, then it is very likely that 
the sum $\sum_{i=1}^m \cos(2 \pi k z_i/N)$ is larger than $m/4$
if and only if $k=k_0$ or $k=N-k_0$.

\begin{proof}\pftext{of Theorem~\ref{thm:peak}}
This is a simple consequence of Lemma~\ref{lm:singlepoint}.
Let $k'_0 = \min\{k_0,N-k_0\}$.
The probability that $\sum_{i=1}^m \cos(2 \pi k'_0 z_i/N) \leq m/4$
is at most $e^{-m/32} \leq \frac{1}{N^2}$. 
Furthermore, for all integers $0 < k \leq N/2$ not equal to~$k'_0$, 
the probability that $\sum_{i=1}^m \cos(2 \pi k z_i/N) \geq m/4$
is also at most $\frac{1}{N^2}$. 
If ${\tilde k} \neq k'_0$, then one of these $\lfloor N/2 \rfloor$ 
events must have happened, and the probability for that 
is upper bounded 
by~$\big\lfloor\frac{N}{2}\big\rfloor \frac{1}{N^2} \leq \frac1{2N}$.
\end{proof}

\section{Abelian Hidden Subgroups}\label{sec:abelian}
Theorem~\ref{thm:abelian} in Section~\ref{sec:alg} states
that the Abelian subgroup problem can be solved efficiently on a quantum
computer.  The algorithm which accomplishes this is most easily
understood using some basic representation theory for finite Abelian
groups which we now briefly review.  For more details see the
excellent references~\cite{MR96,Rockmore96}.  
For any Abelian group~$G$ the
group algebra $\Co[G]$ is the Hilbert space of all complex-valued
functions on $G$ equipped with the standard inner product. 
A~{\em character\/} of~$G$ 
is a homomorphism from $G$ to~$\Co$.  The set of characters
admits a natural group structure via pointwise multiplication and is a
basis for the group algebra.  The {\em Fourier transform\/} is the linear
transformation from the point mass basis of the group algebra to the
basis of characters.  It is known that the quantum Fourier transform
may be performed in time $O\big(\log^2 |G|\big)$.  Finally, for any
subgroup $H \subgroup G$, there exists a subgroup of the character
group called the orthogonal subgroup $H^\perp$ which consists of all 
characters $\chi$ such that $\chi(h) = 1$ for all $h \in H.$  

We~now sketch the quantum algorithm for solving the Abelian hidden
subgroup problem.  In~the interest of clarity 
we omit all normalization factors in our description.
The state of the computer is initialized in the superposition
\[\sum_{g \in G} \ket{g}\ket{\func(g)}.\]
We~then observe the second register with outcome, say,~\mbox{$r \in R$}.
This action serves to place the first register 
into a superposition of all elements that map to~$r$ under~$\func$.  
Because $\func$ is constant and distinct on cosets of~$H$
we may write the state of the computer as
\[\sum_{h \in H}\ket{s + h}\ket{r}\]
for some coset $s + H$ chosen by the observation of the second register.  
Since we will not use the second register or
its contents in the remainder of the algorithm, we express the
state of the computer as a function of the contents of the first
register only, $\sum_{h \in H}\ket{s + h}$.
We~then apply the quantum Fourier transform which results in the state
\[\sum_{h^\prime \in H^\perp} \inner{h^\prime}{s} \;\ket{h^\prime},\]
which may be verified by direct calculation.
% \[\sum_{h^\prime \in H^\perp} \inner{h^\prime}{s} \;\ket{h^\prime}.\]
% This may be verified by direct calculation but is perhaps most easily 
% understood as a direct result of the following basic theorem.
%
% \begin{theorem}
% If $\func: G \rightarrow R$ is constant 
% (but not necessary distinct) on cosets of~$H$,
% then the Fourier transformed of~$\func$
% is nonzero only on elements of~$H^\perp$.
% \end{theorem}
%
Finally, we observe the first register.  Notice that this results in a
uniformly random sample from~$H^\perp$.  

It~can easily be shown that by repeating this experiment 
of order $\log |H^\perp|$ times,
we find a generating set for~$H^\perp$.  
The hidden subgroup~$H \subgroup G$ can then be
calculated efficiently from~$H^\perp$ on a classical computer, 
essentially by linear algebra.
In~summary, the sole purpose of the quantum machine in the above
algorithm is to sample uniformly from~$H^\perp$.

\section{\boldmath A~Generalized $H^\perp$}\label{sec:perp}
We~now briefly discuss the main ideas of harmonic analysis on groups,
stating as facts the main results that we require.
For more detailed information see the excellent 
references~\cite{MR96,Rockmore96}.
Let~$G$ be a (possibly noncommutative) finite group.  
A~representation of~$G$ is a homomorphism
$\rho:~G \rightarrow GL(V_\rho)$ where $V_\rho$ is called the 
{\em representation space\/} of the representation.  
The dimension of~$V_\rho$, denoted~$d_\rho$, 
is called the dimension of the representation.  
The representation~$\rho$ is {\em irreducible\/} 
if the only invariant subspaces of~$V_\rho$ are $0$ and~$V_\rho$ itself.  
Two representations $\rho_1$ and~$\rho_2$ are {\em equivalent\/} 
if there exists an invertible linear map 
$S:V_{\rho_1} \rightarrow V_{\rho_2}$ such that
$\rho_1(g) = S^{-1} \smallspace \rho_2(g) \smallspace S$ 
for all \mbox{$g \in G$}.

Let $\Gamma = \{\rho_1,\rho_2,\dots,\rho_r\}$ be a
complete set of inequivalent, irreducible representations of~$G$. 
Then the identity $\sum_{i=1}^r d_{\rho_i}^{\smallspace 2} = |G|$ holds.
Furthermore, we may assume that the representations are unitary, 
i.e., that $\rho(g)$ is a unitary matrix 
for all $g \in G$ and all~$\rho \in \Gamma$.  
The functions defined by \mbox{$\rho_{ij} = \rho(g)_{ij}$} 
for $1 \leq i,j \leq d_{\rho}$ 
are called {\em matrix coefficients}, and by the previous
identity it follows that there are $|G|$~matrix coefficients.  
It~is a fundamental fact that the set of all {\em normalized\/} matrix
coefficients obtained from any fixed $\Gamma$ is an orthonormal basis
of the group algebra~\mbox{$\Co[G]$}.  
The {\em Fourier transform\/} (with respect to a chosen~$\Gamma$) 
is a change of basis transformation of the group algebra 
from the basis of point masses to the basis of matrix coefficients.  

If~$G$ is commutative, then these definitions reduce to those 
discussed in the previous section, since in that case, 
all representations are 1-dimensional 
and each matrix coefficient is just a character.  
If~$G$ is noncommutative, then there exists at least 
1~irreducible representation of $G$ with higher dimension,
and in this case the Fourier transform depends on the
choice of bases for the irreducible representations.  
It~seems as though this is what complicates 
the extension of the quantum algorithm 
for commutative groups to the noncommutative scenario.

It~turns out that for our present application it is most useful to use
an equivalent notion of the Fourier transform.  One may also think of
the matrix coefficients as collected together in matrices.  In~this
view the Fourier transform is a matrix-valued function on~$\Gamma$.
For each \mbox{$f \in \Co[G]$}, 
we define the value of the Fourier transform at an
irreducible representation~$\rho \in \Gamma$ to~be
\[\hat f(\rho) = \sqrt{\frac{d_\rho}{|G|}\tinyspace} 
  \;\sum_{g \in G}f(g)\rho(g).\]
If~we take individual entries of these matrices, then we recover the
coefficients in the basis of matrix coefficients.  There is a Fourier
inversion formula and therefore $f$ is determined by the matrices
$\big\{\hat f(\rho)\big\}_{\raisebox{2mm}{$\rho \in \Gamma$}}\smallspace$.

We~may now describe the noncommutative version of~$H^\perp$.  
Let~$V_\rho^H$ be the elements of~$V_\rho$ 
that are {\em pointwise\/} fixed by~$H$,
\[V_\rho^H \smallspace=\smallspace 
    \{v \in V_\rho \mid \rho(h)v = v, \;h \in H\}.\]
Let $P_\rho^H$ be the projection operator onto~$V_\rho^H$.  
Then define 
\[H^\perp = 
  \big\{P_\rho^H\big\}_{\raisebox{2mm}{$\rho \in \Gamma$}}\smallspace.\]
The significance of this definition follows from the 
following elementary \mbox{result.}
\begin{theorem}
Let~$I_H$ be the indicator function on the subgroup~$H \subgroup G$.  
Then, for all $\rho \in \Gamma$, we have that $\hat{I}_H(\rho) = P_\rho^H$. 
\end{theorem}

\begin{corollary}
Let $sH$ be any coset of~$H \subgroup G$.
Then the previous theorem immediately yields
$\hat{I}_{sH}(\rho) = \rho(s)P_\rho^H$.
\end{corollary}

Let us briefly summarize the role of this result in the quantum
algorithm.  If~we straight-forwardly apply the quantum algorithm
described in the previous section to the case where $G$ is
noncommutative, then we must determine the resulting probability amplitudes
and the information gained by sampling according to these amplitudes.

Recall that the state of the quantum system 
after the first observation is a superposition 
of states corresponding to the members of one coset.  
Thus the state may be described by the indicator function
of a coset~$I_{sH}$.  
The final observation results in observing the 
name of a matrix coefficient~\ket{\rho,i,j}.  
The probability of observing~\ket{\rho,i,j} 
is given by~$|c_{\rho,i,j}|^2$ 
where $c_{\rho,i,j}$ is the coefficient of~$\rho_{ij}$ 
in the expansion of~$I_{sH}$ in the basis of matrix coefficients.  
The corollary above allows us, in~theory, 
to compute these probability amplitudes.

The algorithm described in the first part of this paper may be derived
from these general methods.  For a general noncommutative group it
seems that these methods are necessary for an analysis of the
resulting probability amplitudes.

\section{Acknowledgements}
We~would like to thank Dan Rockmore, David Maslen and 
Hans~J.{} Munkholm from whom we learned noncommutative Fourier analysis,
and Richard Hughes, Robert Beals and Joan Boyar for helpful conversations 
on this problem.

\end{document}